\documentclass[preprint,12pt,authoryear]{elsarticle}

\usepackage{cite}
\usepackage{amssymb}

\usepackage{amsmath}

\usepackage{booktabs}
\usepackage{graphicx} 
\RequirePackage{dsfont}
\RequirePackage{amssymb}
\usepackage{colortbl,color,soul}
\usepackage{algorithm}
\usepackage{eurosym}
\usepackage{algpseudocode}
\usepackage{nicefrac}
\RequirePackage[usenames,dvipsnames,table]{xcolor}
\usepackage[%
  colorlinks = true,
  citecolor  = Black,
  linkcolor  = Black,
  urlcolor   = Black,
  unicode,
  ]{hyperref}
\usepackage{cleveref}
\RequirePackage[%
  acronym,
  automake,
  nogroupskip,
  nopostdot,
  nonumberlist,
  toc,
  ]{glossaries}
\usepackage{mathtools} 
\usepackage{indentfirst}

\newacronym{RES}{RES}{renewable energy sources}
\newacronym{UC}{UC}{unit commitment}
\newacronym{SFR}{SFR}{system frequency response}
\newacronym{ED}{ED}{dconomic dispatch}
\newacronym{UFLS}{UFLS}{under frequency load shedding}
\newacronym{RoCoF}{RoCoF}{rate of change of frequency}
\newacronym{KDE}{KDE}{kernel density estimate}
\newacronym{MILP}{MILP}{mixed integer linear programming}
\newacronym{MAE}{MAE}{mean absolute error}
\newacronym{MSE}{MSE}{mean squared error}
\newacronym{ESS}{ESS}{energy storage system}
\newacronym{ANN}{ANN}{artificial neural network}
\newacronym{OCT}{OCT}{optimal classification tree}
\newacronym{DNN}{DNN}{deep neural network}
\newacronym{ML}{ML}{machine learning}

\makeglossaries

\journal{Engineering Applications of Artificial Intelligence}

\begin{document}

\begin{frontmatter}

\title{Data-driven Estimation of Under Frequency Load Shedding after Outages in Small Power Systems} 

\author[inst1]{Mohammad Rajabdorri} 
\author[inst2]{Matthias C. M. Troffaes}
\author[inst2]{Behzad Kazemtabrizi}
\author[inst1]{Miad Sarvarizadeh-Kouhpaye}
\author[inst1]{Lukas Sigrist}
\author[inst1]{Enrique Lobato}
\affiliation[inst1]{organization={IIT, Comillas Pontifical University},
            country={Spain}}
            
\affiliation[inst2]{organization={Durham University},
            country={UK}}

\begin{abstract}
This paper presents a data-driven methodology for estimating \gls{UFLS} in small power systems. \Gls{UFLS} plays a vital role in maintaining system stability by shedding load when the frequency drops below a specified threshold following loss of generation. Using a dynamic \gls{SFR} model we generate different values of \gls{UFLS} (i.e., labels) predicated on a set of carefully selected operating conditions (i.e., features). \Gls{ML} algorithms are then applied to learn the relationship between chosen features and the \gls{UFLS} labels. A novel regression tree and the Tobit model are suggested for this purpose and we show how the resulting non-linear model can be directly incorporated into a \gls{MILP} problem. The trained model can be used to estimate \gls{UFLS} in security-constrained operational planning problems, improving frequency response, optimizing reserve allocation, and reducing costs. The methodology is applied to the La Palma island power system, demonstrating its accuracy and effectiveness. The results confirm that the amount of \gls{UFLS} can be estimated with the \gls{MAE} as small as 0.213 MW for the whole process, with a model that is representable as a \gls{MILP} for use in scheduling problems such as unit commitment among others.
\end{abstract}


\begin{highlights}
\item The paper introduces a learning process to estimate the amount of \gls{UFLS}, focusing on conventional schemes due to their widespread use.
\item Dataset generation, feature selection, and learning models are carefully analyzed, and a novel partitioning algorithm for regression trees is proposed.
\item The paper provides a careful analysis of \gls{UFLS} estimation and presents \gls{MILP} representations of \gls{UFLS} estimation, which is novel in the literature.
\end{highlights}


\end{frontmatter}

\printglossary[type=\acronymtype]

\section{Introduction}
\label{sec:intro}
Synchronous generators are being displaced with cleaner, albeit non-synchronously coupled alternatives (like wind and solar), which inadvertently has led to a reduction of inertia, hence incorporating frequency dynamics into the operational planning of power systems is more important than ever. Island power systems are already suffering from a lack of inertia because of their small size. There has been extensive research on how to include frequency dynamics in scheduling optimization problems. Both analytical methods (directly from the swing equation) and data-driven methods (based on dynamic simulations) have been proposed to obtain frequency constraints for inclusion in the operational planning process. Typical frequency response metrics after outages are the \gls{RoCoF}, quasi-steady-state frequency, and frequency nadir. However, calculating the frequency nadir is much more complicated than the other metrics. To derive the frequency nadir from the swing equation, some simplifying assumptions are needed, and still, the obtained equation is non-linear and non-convex, which makes it challenging to be used in \gls{MILP} problem formulations. A common assumption in analytical frequency-constrained methods like \cite{trovato2018unit, badesa2019simultaneous, paturet2020stochastic, shahidehpour2021two, ferrandon2022inclusion}, and many other similar works, is that the provision of reserve increases linearly in time, and all units will deliver their available reserve within a given fixed time. Consequently, the ensuing complicated analytical methods are not necessarily accurate. On the other hand, more recently, \gls{ML}-based methods have been proposed to incorporate frequency dynamics. For instance, optimal classifier tree is used in \cite{lagos2021data}, deep neural network is used in \cite{zhang2021encoding}, and logistic regression is used in \cite{rajabdorri2022robust}, among other approaches. An analytical frequency constrained \gls{UC} is compared with data-driven models with the help of \gls{ML} in \cite{rajabdorri2023inclusion} and their pros and cons are highlighted.

In smaller systems like islands the frequency can easily exceed the safe threshold after any contingency because usually online units are providing a considerable percentage of the whole demand. To maintain the frequency stability of an electrical power system, \gls{UFLS} schemes are implemented to shed or disconnect a certain amount of load in predefined steps when the frequency drops below a specified threshold following disturbance events. This corrective protection measure helps to balance the power supply and demand and prevents a complete system blackout. Different methods have been introduced to tune and optimize the \gls{UFLS} scheme for electrical power systems, which can be categorized into conventional and adaptive methods. Conventional \gls{UFLS} schemes use fixed load shedding steps \cite{ketabi2014underfrequency, laghari2014new, wang2022underfrequency, kalajahi2021under}, while adaptive \gls{UFLS} schemes dynamically adjust the load shedding amount based on real-time system conditions \cite{mehrabi2018toward, li2019continuous, tofis2016minimal, silva2020adaptive}. Although adaptive schemes provide a more optimized and flexible response, as they require more advanced monitoring systems and computational capabilities for real-time monitoring and optimization, they are still not used widely in practice. The performance of both conventional and adaptive methods can be improved by incorporating \gls{ML} \cite{hooshmand2012optimal, golpira2022data}.

Depending on the size of the system, it is possible to prevent \gls{UFLS} activation. Many studies like \cite{zhang2021encoding, chang2012frequency, sedighizadeh2019optimal, perez2016robust}, and others set the frequency dynamic thresholds high enough, so no outage leads to \gls{UFLS}. This is not possible in a small system, where every online unit is providing a substantial percentage of the demand and any outage can be big enough to trigger the \gls{UFLS} activation \cite{rajabdorri2022robust}. In such systems co-optimizing \gls{UFLS} activation and scheduling of the units can have some benefits like:
\begin{itemize}
    \item Having an estimate of \gls{UFLS} in the scheduling optimization problem will prevent incidents with poor frequency responses.
    \item The estimated amount of \gls{UFLS} can be deduced from the required reserve. There is no need to schedule reserve as much as the biggest outage if eventually the \gls{UFLS} will be activated after the outage.
    \item \Gls{UFLS} can be monetized easily and added directly to the objective function of the optimization problem, to reduce the overall operation costs.
\end{itemize}

This paper tries to estimate the amount of \gls{UFLS}, regardless of the \gls{UFLS} scheme that is used for the system, through a learning process. The estimation of the amount of \gls{UFLS} of conventional schemes is however very complicated because of its discrete nature and disturbance-dependent behavior. Given the widespread use of conventional schemes in real systems, it's the principal focus of this paper.
The dataset used for the learning process is labeled with the \gls{UFLS} of every sample generation combination in the dataset. The labels can be obtained by \gls{SFR} models or any other power system simulator.
The purpose is to use the estimation of \gls{UFLS} in the operational planning of small power systems (such as generation \gls{UC}, \gls{ED}, reserve allocation, ancillary service scheduling, \gls{RES} integration, and so on). The operational planning process is usually modeled and solved as an \gls{MILP} problem. Therefore, it is convenient to limit the hypothesis space of the \gls{UFLS} estimation models to models that are representable by \gls{MILP}. Including \gls{UFLS} estimation in the problem is in a sense equivalent to including frequency dynamics in the operational planning, because poor frequency response subsequently triggers the activation of \gls{UFLS}. 

To estimate the amount of \gls{UFLS}, a dataset generation process is proposed to acquire a set of operating points (potential hourly generation schedules) that can properly describe the system under study. Every possible outage in each set of generation schedules is labeled with its corresponding \gls{UFLS}. The obtained dataset is carefully analyzed to choose the representative features and then a learning process is proposed. A regression tree model with a novel partitioning algorithm is suggested in this paper. As the \gls{UFLS} is activated in steps and discretely sheds load, a regression tree seems most suitable. Our partitioning algorithm exploits the data structure to most effectively represent the regression tree as a \gls{MILP} \cite{maragno2021mixed}. Also, the use of the Tobit model to estimate \gls{UFLS} is proposed and studied. Although the Tobit model is typically used to describe censored data \cite{tobin1958estimation}, we found it to be effective to also describe the \gls{UFLS}, which has a cluster around zero followed by a linearly increasing trend as a function of the features, very similar to zero censored datasets. Additionally, the Tobit model has a simple analytic structure that makes it easy to incorporate into a \gls{MILP}. Both the suggested regression tree and the Tobit model, alongside their \gls{MILP} representation are demonstrated in the following. To the best of the author's knowledge, there's no prior careful analysis on \gls{UFLS} estimation or an \gls{MILP} representation of \gls{UFLS} in the literature.

The rest of the paper is organized as follows; in \cref{sec:metho} the methodology of the paper is introduced, including the data generation (in \cref{sec:data_gen}), labeling the data (in \cref{sec:labelling}), and the learning process (in \cref{sec:learning}). Then in \cref{sec:results} the results for the island under study (La Palma) are presented, including the data generation and analysis (in \cref{sec:data_analysis}), the applied learning process, and its accuracy (in \cref{sec:learning_res}). Finally, conclusions are drawn in \cref{sec:conc}.

\section{Methodology}\label{sec:metho}

\subsection{Data Generation}\label{sec:data_gen}

To estimate the \gls{UFLS}, a proper set of data is necessary. The training dataset comprises features $x \in \mathcal{X}$ and labels $y \in \mathcal{Y}$. In the case of implementing the estimation of \gls{UFLS} in the scheduling problems (like \gls{UC}), features are any measurable quantities from the power system that might help predict the \gls{UFLS}. These features are extracted from generation combinations, while the labels are obtained from dynamic simulations of \gls{UFLS} after outages. These measurements can be obtained by solving high-order differential swing equations, or by using \gls{SFR} models. The features should be carefully chosen to represent a reasonable amount of information about their labels. Using an unnecessarily large number of features can be detrimental to both computational and statistical aspects. Selecting a large feature vector increases the dimensions of the problem, thereby requiring more resources for calculations. In addition, using a higher number of features makes the model more susceptible to overfitting. Therefore, it is beneficial to use only the features with the most relevant information to predict the label $y$ \cite{Jung2018}. In this paper, $y$ is the amount of \gls{UFLS} for each outage. Several methods have been introduced in the literature to reduce the size of the feature vector. For this paper, the features must be accessible throughout the scheduling process. Therefore, variables that are most correlated with the label will be chosen as features. As shown later in \cref{sec:results}, the selected features for predicting \gls{UFLS} are available inertia ($\mathcal{H}_g$), weighted gain of turbine-governor model ($\mathcal{K}_g$), the amount of lost power ($P_g$), and the amount of available reserve ($\mathcal{R}_g$), after the outage of generator $g$.

In order to estimate the \gls{UFLS}, a proper set of simulation data is necessary. The training dataset comprises features $x \in \mathcal{X}$ and labels $y \in \mathcal{Y}$. In the case of implementing the estimation of \gls{UFLS} in the operational planning problems (like \gls{UC}), features are any measurable quantities from the power system that might help predict the \gls{UFLS}. These features are extracted from generation combinations, while the labels are obtained from dynamic simulations of \gls{UFLS} after outages. It is beneficial to use only the features with the most relevant information to predict the label $y$ \cite{Jung2018}. In this paper, $y$ is the amount of \gls{UFLS} following each outage, and variables that are most correlated with the label are chosen as features. As shown later in \cref{sec:results}, the selected features for predicting \gls{UFLS} are available inertia ($\mathcal{H}_g$), the weighted gain of turbine-governor model ($\mathcal{K}_g$), the amount of lost power ($P_g$), and the amount of available reserve ($\mathcal{R}_g$), after the outage of generator $g$. The synthetic data generation method introduced in \cite{rajabdorri2023inclusion}, is used here. The obtained dataset consists of all feasible and economical generation combinations in La Palma Island, which is the case study of this paper.

To obtain a complete dataset, every combination of possible generation outputs of the units can be considered. However, many of these combinations are infeasible as they do not satisfy the constraints that are used in the scheduling process (power balance, reserve constraint, or maximum \gls{RoCoF}), or are unappealing as the optimization problem will favor cheaper combinations. In this paper, a data generation method is used to only generate feasible control points that are cost-effective, and hence more likely to be scheduled in the real operation. The process is outlined in \cref{algori}.

\begin{algorithm}[!htbp]
\caption{Synthetic Data Generation}
\label{algori}
\textbf{Inputs}:
$\overline{D}$, $\underline{D}$, $\overline{P}_i$, $\Delta f$, $f_0$, $H_i$, $M_i$\\
\textbf{Output}: $\mathcal{F}$: set of feasible power level vectors
\begin{algorithmic}[1]
    \State $\mathcal{F}\gets\emptyset$
    \For{$\vec{p}\in \bigtimes_{i=1}^I \mathcal{P}_i$} \Comment{\texttt{\small for all power level vectors}}
    \For{$i\in\{1,\dots,I\}$}\Comment{\texttt{\small for every generator}}
    \State $u_i\coloneqq 0\text{ if }p_i=0\text{ else }1$ \Comment{\texttt{\small status of unit}}
    \EndFor
    \State $G\coloneqq\sum_{i=1}^I p_i$ \Comment{\texttt{\small total generation}}
    \State $R_\ell\coloneqq \big(\sum_{i=1}^I u_i(\overline{P}_i-p_i)\big)-u_\ell(\overline{P}_\ell-p_\ell)$ \Comment{\texttt{\small reserve after outage of $\ell$}}
    \State $H_\ell^{\text{sys}}\coloneqq(\sum_{i=1}^I H_i M_i u_i)-H_\ell M_\ell u_\ell$\Comment{\texttt{\small inertia after outage of $\ell$}}
    \If{$\underline{D}\le G\le \overline{D}$ and $R_\ell\ge p_\ell$ and $H_\ell^{\text{sys}}\geq \frac{p_\ell f_0}{2\Delta f}$}
    \Comment{\texttt{\small feasible?}}
    \State $\mathcal{F}\gets \mathcal{F}\cup \{\vec{p}\}$ \Comment{\texttt{\small add power level vector}}
    \EndIf
    \EndFor
    \State Sort $FC$ ascending by the quadratic generation cost function
    \State Keep a reasonable number of cheaper combinations and remove the rest
\end{algorithmic}

\hrulefill

\small $\overline{D}$ and $\underline{D}$ are lower and upper bounds on yearly thermal generation (MW), $i$ is the index of unit $\in\{1,\dots,I\}$, $\overline{P}_i$ is the capacity of unit $i$ (MW), $\Delta f$ is critical \gls{RoCoF} (Hz/s), $f_0$ is nominal frequency (Hz), $\mathcal{P}_i$ is the finite set of power levels of unit $i$ including level $0$ for not committed (MW), $\ell$ is the index of the lost unit (can be any $i$), $H_i$ is the inertia of unit $i$, and
$M_i$ is the base power of unit $i$.

\end{algorithm}

\subsection{Labeling the Data}\label{sec:labelling}

\begin{figure}[!htbp]
\centering
\includegraphics[width=0.65\linewidth]{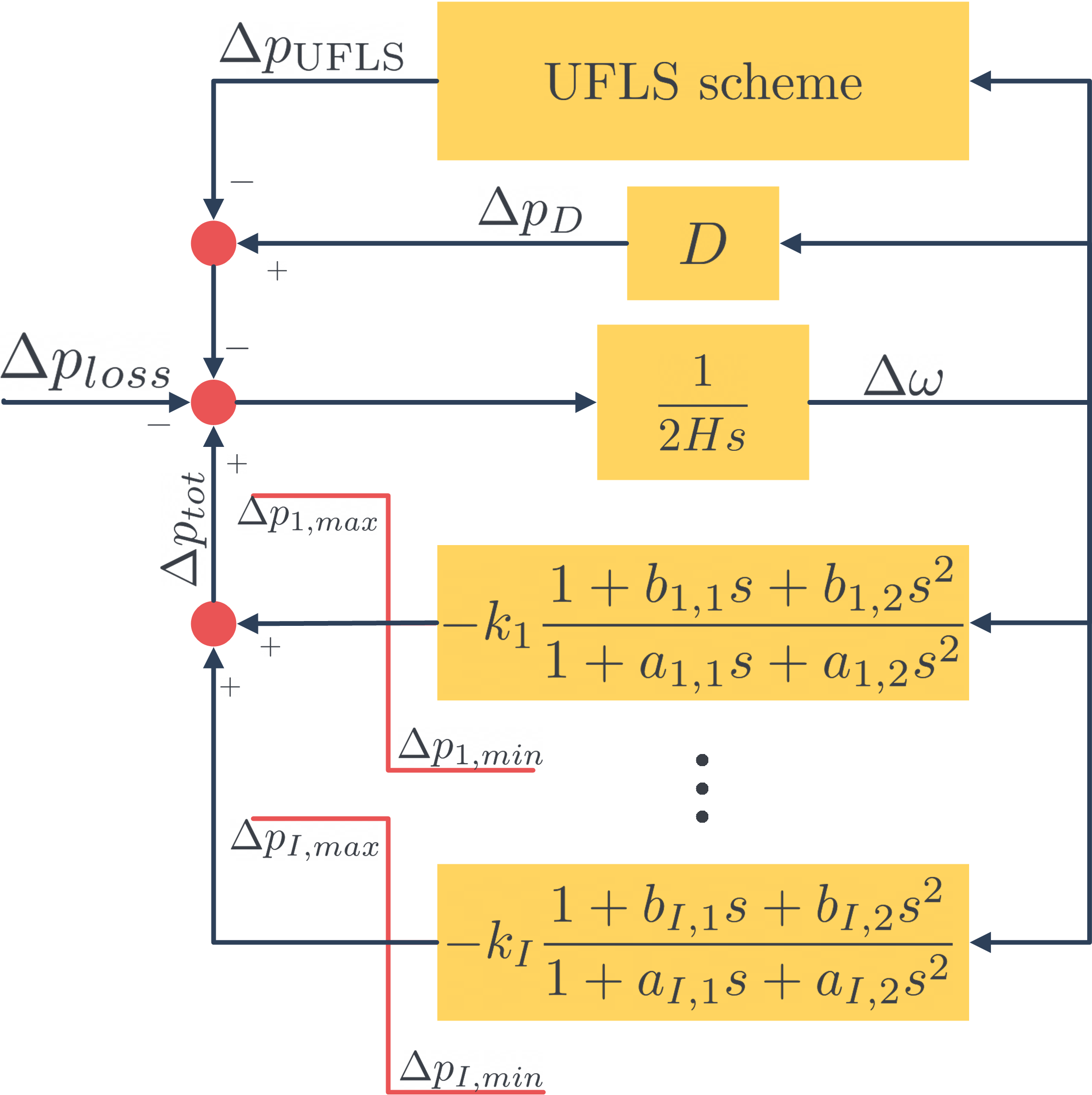}
\caption{\gls{SFR} model.}
\label{fSFRmodel}\end{figure}

In labeling the data, the \gls{SFR} model is used to analyze the frequency stability of small isolated power systems, such as the La Palma Island system being studied. The \gls{SFR} model can reflect the short-term frequency response of such systems, but other dynamic power system models could also be employed. The power-system model, which is typically used to design \gls{UFLS} schemes for an island power system consisting of $I$ generating units, is detailed in \cref{fSFRmodel}.
A second-order model approximation is used to represent the turbine-governor system of each generating unit ($i$). The dynamic frequency responses are mainly influenced by the rotor and turbine-governor system dynamics, while excitation and generator transients are ignored due to their faster dynamics. The load-damping factor ($D$) is used to consider the overall response of the loads, provided that its value is known. The gain ($k_i$), which is the inverse of the droop, and the parameters ($a_{i,1}$, $a_{i,2}$, $b_{i,1}$, and $b_{i,2}$) of each generating unit ($i$) can be determined from more precise models or field tests. The gain ($k_i$) is an essential parameter to indicate the frequency response of unit $i$, and will influence the \gls{UFLS} scheme activation. To have features that can reflect the amount of \gls{UFLS} after the outage, a weighted gain is defined, which will be used as a feature for the training dataset. The equation,
\begin{equation}
    \mathcal{K}_g=\sum\limits_{i\in \mathcal{I}, i\neq g} k_i\mathcal{M}_i u_{t,i}
\end{equation}
represents the weighted gain after the outage of unit $g$. Due to the limited primary spinning reserve, the units' power output is restricted by the power output limitations $\Delta p_{i,min}$ and $\Delta p_{i,max}$, and the ramp-up speed of the units should be constrained by the maximum ramping capacity of each respective unit. The complete model is explained in \cite{Sigrist2016}.

\subsection{Learning Process}\label{sec:learning}

\subsubsection{Proposed Regression Tree}

A regression tree is suggested in this paper to estimate the amount of \gls{UFLS} since conventional schemes shed load in a discrete manner. Typical regression trees split the feature space into rectangular cells and use a constant value within each cell for prediction.
However, if we want to incorporate the model into a \gls{MILP}, we prefer to have as few cells as possible.
Therefore, a novel regression tree is proposed here, which is inspired by \cite{verwer2017auction} but deviates from it in several ways:
\begin{itemize}
\item The data shows that using convex regions leads to a far more efficient representation. Therefore, we use linear functions to partition cells, rather than single features.
\item Within each cell, a linear model is used instead of a constant, as this further reduces the number of cells.
\item For prediction, it's important to estimate incidents with no \gls{UFLS} as exactly zero and not a small number. The suggested tree structure can achieve that.
\end{itemize}
Representing such regression tree as \gls{MILP} is presented later. \gls{MILP}-representability of many \gls{ML} methods is exploited in \cite{maragno2021mixed}.

\begin{figure}[!htbp]
\centering
\includegraphics[width=0.8\linewidth]{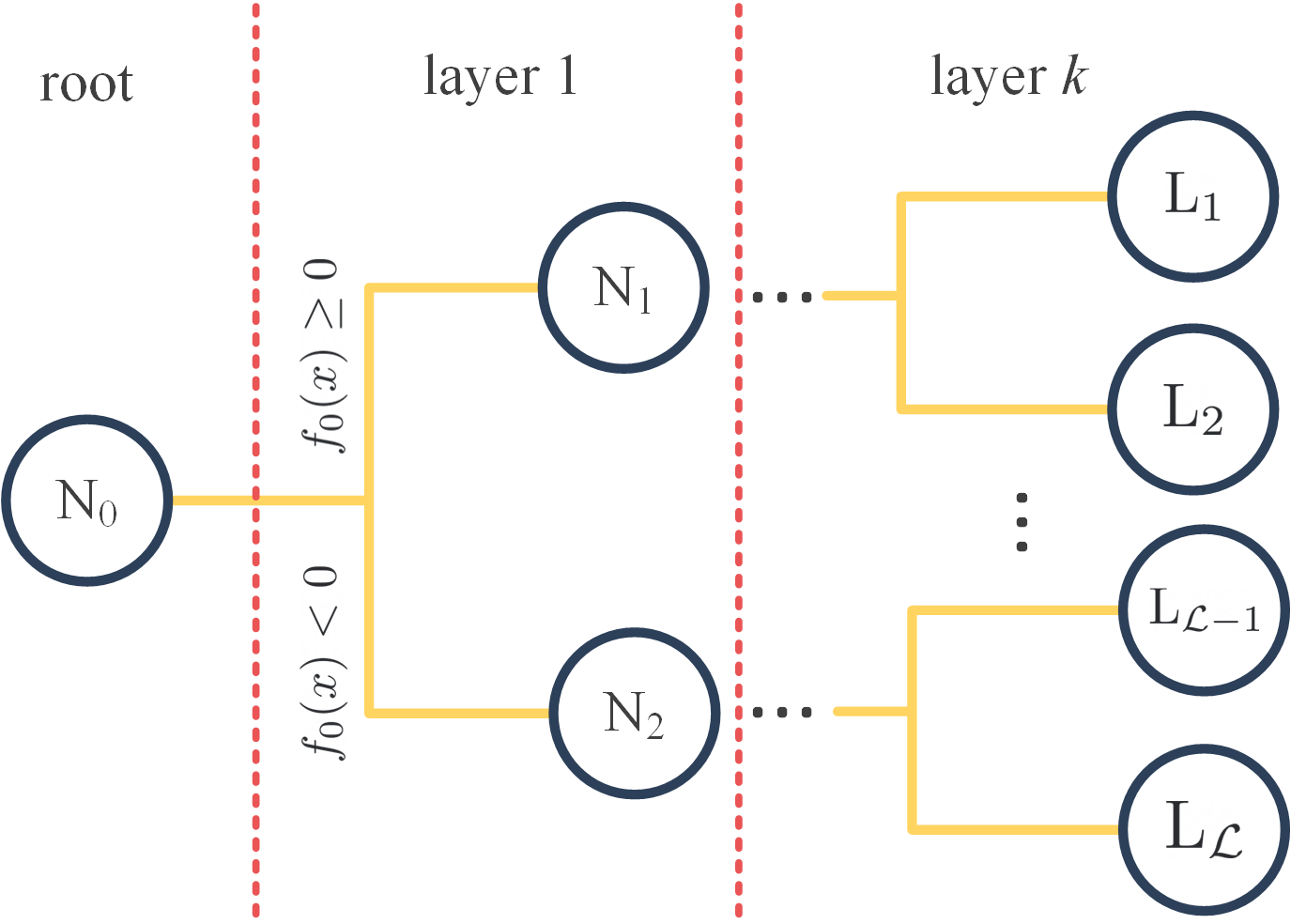}
\caption{Proposed regression tree.}
\label{fig:binary_tree}\end{figure}

The suggested regression tree is shown in \cref{fig:binary_tree}. In this figure, N$_0$ is the root node. N$_1$ and N$_2$ are the nodes of the first layer. A linear function of the features (for example $f_0(x)$ for the root node) will split the nodes into two to classify the incidents with a threshold on the labels. Then on the last layer, there are the leaves L$_1$ to L$_\mathcal{L}$. Linear regression is applied to the samples within each of these leaves.
As the labels of the dataset are already accessible after the labeling process (\cref{sec:labelling}) different methods can be applied to split each node. Each splitting function is found by solving a univariate optimization problem, as explained next. A grid search is performed to find the optimal cut-point for each split ($c$ in \cref{fig:binary_tree}). Splitting nodes is continued until the \gls{MAE} of the new structure is higher than before splitting. From the results that are obtained from the grid search, the one with the best overall \gls{MAE} will be picked.  

Finding the optimal linear function $f(x)=\beta_0+\sum_{i=1}^p\beta_i x_i$, where $p$ is the number of features, to split each cell requires solving a $p+1$ dimensional optimization problem, which finds the coefficients $\beta$ that minimizes the error of the local linear fits in each cell.
This is a difficult optimization problem, and also a reason why typical regression trees split only on single features, i.e. only on functions of the form $f(x)=x_i-c$ for a single feature $x_i$.
To reduce the problem of finding $\beta$ to a univariate optimization problem, we propose a model inspired by \cite{verwer2017auction}, but for splitting the nodes logistic regression is used instead of assigning a threshold to a feature and instead of assigning a constant at the leaf nodes a linear regression is applied. Considering that the \gls{UFLS} from the \gls{SFR} model is either zero or a positive number, it's important to estimate the incident with no \gls{UFLS} as zero and not a small number. This also can be achieved with this tree structure.

We first introduce a univariate threshold variable $c$ (over which we will optimize), and binary variable $z$, with $z=0$ if $y< c$ (here, $y$ is the amount of \gls{UFLS}, i.e. the value we try to predict) and $z=1$ otherwise. Define the logistic function as follows:
\begin{equation}
p_\beta(x)\coloneqq \frac{1}{1+\exp(-\beta_0-\sum_{i=1}^p \beta_i x_i)}
\end{equation}
The key idea now is that a good split for predicting $y$ should be able to predict $z$ from $x$. Therefore, we find $\beta_0$, \dots, $\beta_p$ to maximize the log-likelihood,
\begin{equation}
\sum_{j\in N} z^{(j)}\log p_\beta(x^{(j)})+(1-z^{(j)})\log (1-p_\beta(x^{(j)}))
\end{equation}
where $N$ is the node (as a subset of sample indices) that we are currently trying to split.
For any given threshold $c$, denote the maximum likelihood estimate of $\beta$ by $\hat{\beta}(c)$. 
Consequently, for each $c$, we can split node $N$ into two sub-nodes $N'(c)$ and $N''(c)$. Now by performing a one-dimensional grid search, we choose $c$ that minimizes the overall error of the local linear models. The splitting can be continued until the tree can predict the amount of \gls{UFLS} with an acceptable accuracy. Note that a simpler tree structure is preferred for two reasons. Firstly, although the accuracy of the model for the dataset in hand might improve by adding more layers, the model will be more susceptible to over-fitting. Secondly, the \gls{MILP} representation becomes computationally burdensome for more complicated tree structures.

To predict values of the \gls{UFLS} on each leaf, standard linear regression is used. The maximum likelihood estimate of the parameters of the linear model of leaf L$_\ell$ can be calculated by finding $\alpha_0,\dots,\alpha_p$ that minimize
\begin{equation}
\sum_{\substack{j\in \text{L}_\ell}}
\left(\alpha_0+\sum_{i=1}^p \alpha_i x_i^{(j)}-y^{(j)}\right)^2
\end{equation}
Note that, for this fit, only the samples that are assigned to the leaf L$_\ell$ of the \gls{UFLS} data are included, resulting in the summation over indices $j\in \text{L}_\ell$.

Putting it all together,
we predict $\hat{y}$ from $x$ using the following piecewise linear function,
\begin{equation}
\hat{y}(x)=
\begin{cases}
\alpha_0^{1}+\sum_{i=1}^p \alpha_i^{1} x_i & \text{if } x\in \text{L}_1 \\
\vdots\\
\alpha_0^{\ell}+\sum_{i=1}^p \alpha_i^{\ell} x_i & \text{if } x\in \text{L}_\ell\\
\vdots\\
\alpha_0^{\mathcal{L}}+\sum_{i=1}^p \alpha_i^{\mathcal{L}} x_i & \text{if } x\in \text{L}_\mathcal{L}
\end{cases}\label{eq:y_hat}
\end{equation}

\textbf{Encoding the proposed regression tree as \gls{MILP}:} To encode the regression tree (i.e. \cref{eq:y_hat})
as an \gls{MILP} model, first a binary variable $u_\ell$ for each leaf L$_\ell$ needs to be defined, which is equal to 1 if $x$ belongs to leaf L$_\ell$.
Since $x$ can only belong to one leaf (see \eqref{eq:y_hat}), the sum of the binary variables $u$ is equal to 1:
\begin{equation} \label{eq:sumu}
    \sum_{\ell\in\mathcal{L}} u_\ell = 1
\end{equation}
Further, the binary variables $u$ should be equal to 0 if any of the parent nodes fails. Finally, the decisions at the non-leaf nodes (N$_0$ to N$_\mathcal{N}$) directly influence the values of the binary variables of the downstream leaves \cite{verwer2017auction}. For instance, if the decision at N$_0$ is $f_0(x)=\hat{\beta}_0^{0}+\sum_{i=1}^p \hat{\beta}_i^{0} x_i < 0$), then $u_\ell=0$ for leaves in the upper subtree, and $u_\ell=1$ for leaves in the lower subtree. The following two constraints force $u_\ell$ to take these values as a function of $x$:
\begin{subequations}
\begin{align}
     \hat{\beta}_0^{0}+\sum_{i=1}^p \hat{\beta}_i^{0} x_i+\underline{\mathcal{M}}_0\sum_{\ell\in \mathcal{L}'} u_\ell &\ge \underline{\mathcal{M}}_0\\
     \hat{\beta}_0^{0}+\sum_{i=1}^p \hat{\beta}_i^{0} x_i+\overline{\mathcal{M}}_0\sum_{\ell\in \mathcal{L}''} u_\ell &< \overline{\mathcal{M}}_0
\end{align}
\end{subequations}
where $\hat{\beta}^{(0)}$ are the obtained logistic regression coefficients for node N$_0$. $\underline{\mathcal{M}_0}$ and $\overline{\mathcal{M}_0}$ are lower and upper bounds for the values that $\hat{\beta}_0^{0}+\sum_{i=1}^p \hat{\beta}_i^{0} x_i$ can take for any $x$ in N$_0$. $\mathcal{L}'$ and $\mathcal{L}''$ are the list of leaves in the upper and lower subtrees of the node N$_0$. A similar set of constraints must be defined for each node.
Now that a binary variable pointing to the correct leaf is accessible, $\hat{y}$ in \cref{eq:y_hat} can be calculated as:
\begin{equation}
    \hat{y}=\sum_{\ell\in\mathcal{L}}u_\ell\times\left(\alpha_0^{\ell}+\sum_{i=1}^p \alpha_i^{\ell} x_i\right)
    \label{eq:y_milp}
\end{equation}
\Cref{eq:y_milp} is non-linear due to the product of binary and continuous variables. To linearize \eqref{eq:y_milp}, the following sets of constraints need to be defined for each leaf $\ell$,
\begin{subequations}
    \begin{align}
        \alpha_0^{\ell}+\sum_{i=1}^p \alpha_i^{\ell} x_i - \overline{\mathcal{M}}_\ell(1-u_\ell)&\leq r_\ell\\
        \alpha_0^{\ell}+\sum_{i=1}^p \alpha_i^{\ell} x_i - \underline{\mathcal{M}}_\ell(1-u_\ell)&\geq r_\ell\\
        \overline{\mathcal{M}}_\ell u_\ell&\geq r_\ell\\
        \underline{\mathcal{M}}_\ell u_\ell&\leq r_\ell
    \end{align}
\end{subequations}
where $\overline{\mathcal{M}}_\ell$ and $\underline{\mathcal{M}}_\ell$ are upper and lower bounds of the term $\alpha_0^{\ell}+\sum_{i=1}^p \alpha_i^{\ell} x_i$ for all $x\in\text{L}_\ell$, and $r_\ell$ is an auxiliary variable. Now the linear equation for $\hat{y}$ can be simply written as,
\begin{equation}
    \hat{y}=\sum_{\ell\in\mathcal{L}} r_\ell
\end{equation}

Considering the proposed \gls{MILP} encoding method, after training the regression tree with $\mathcal{N}$ nodes and $\mathcal{L}$ leaves with the proposed method, to estimate a new observation in a \gls{MILP} model, $2\times\mathcal{N}$ constraints are needed to present the tree structure, one constraint to one-hot encode the binary variables $u$, $4\times\mathcal{L}+1$ constraints to calculate the linearized $\hat{y}$. Also for each observation, $\mathcal{L}$ new continuous variables and $\mathcal{L}$ binary variables are defined.

\subsubsection{Tobit Model}\label{sec:tobit_metho}

Instead of making use of a binary tree and linear regression, we can also consider the standard Tobit model \cite{tobin1958estimation}. The model considers $y^{*}=\alpha_0+\sum_{i=1}^p \alpha_i x_i+\epsilon$, with $\epsilon\sim N(0,\sigma^2)$, but instead of $y^{*}$ the following is observed,
\begin{equation}
    \hat{y}=\begin{cases}
    y^{*} & \text{if $y^{*} > 0$} \\
    0 & \text{otherwise}
  \end{cases}
\end{equation}
The \gls{MILP} representation is straightforward. Same as before, one binary variable is defined for each side.
\begin{subequations}\label{eq:tobit_lin}
\begin{align}
    \alpha_0+\sum_{i=1}^p \alpha_i x_i + \underline{\mathcal{M}}u_{R}&\geq \underline{\mathcal{M}}\\
    \alpha_0+\sum_{i=1}^p \alpha_i x_i + \overline{\mathcal{M}}u_{L}&< \overline{\mathcal{M}}\\
    u_R+u_L&=1\\
    \hat{y}=(\alpha_0+\sum_{i=1}^p \alpha_i x_i)\times u_R&+0\times u_L \label{eq:y_hat_tobit}
\end{align}    
\end{subequations}
As seen in \cref{eq:y_hat_tobit} a variable in binary multiplication appears, which can be linearized as stated before. After training the dataset with the Tobit model, to calculate $\hat{y}$ for each observation in the \gls{MILP} model, 12 constraints, 2 binary variables, and 2 continuous variables are added.

\section{Results}\label{sec:results}

The process of data generation, labeling, data analysis, learning model, and the obtained accuracy for the process are presented in this section. The case study is La Palma Island. The data regarding the island is presented in \cite{rajabdorri2022robust}.

\subsection{Data Generation and Analysis}\label{sec:data_analysis}

The algorithm presented in \cref{algori} is utilized to construct a training dataset for La Palma Island. The power levels are defined using increments of 0.5 MW to form a vector, and all possible combinations of the generators are listed. However, any combinations exceeding the annual thermal generation peak or falling below the annual thermal generation minimum are excluded. The historical thermal generation data for La Palma island indicates that the thermal generation ranges between 36 MW and 16 MW throughout the year. Therefore, the training dataset should only include generation combinations between the maximum and minimum thermal generation limits. Generation combinations that violate the technical requirements are not feasible and are excluded.

The remaining operation points are then sorted by the total value of their quadratic generation cost functions, and the cheaper ones are retained for every thermal generation level. The reason is that these operation points are more likely to appear in the optimized solution. As previously explained, all data points must be labeled with the \gls{SFR} model. The \gls{SFR} model is used to calculate the amount of \gls{UFLS} for each respective outage, with over 110,000 possible outages for the training dataset. The aim is to label all outages with their expected amount of \gls{UFLS}.

\begin{figure}[!htbp]
\centering
\includegraphics[width=\linewidth]{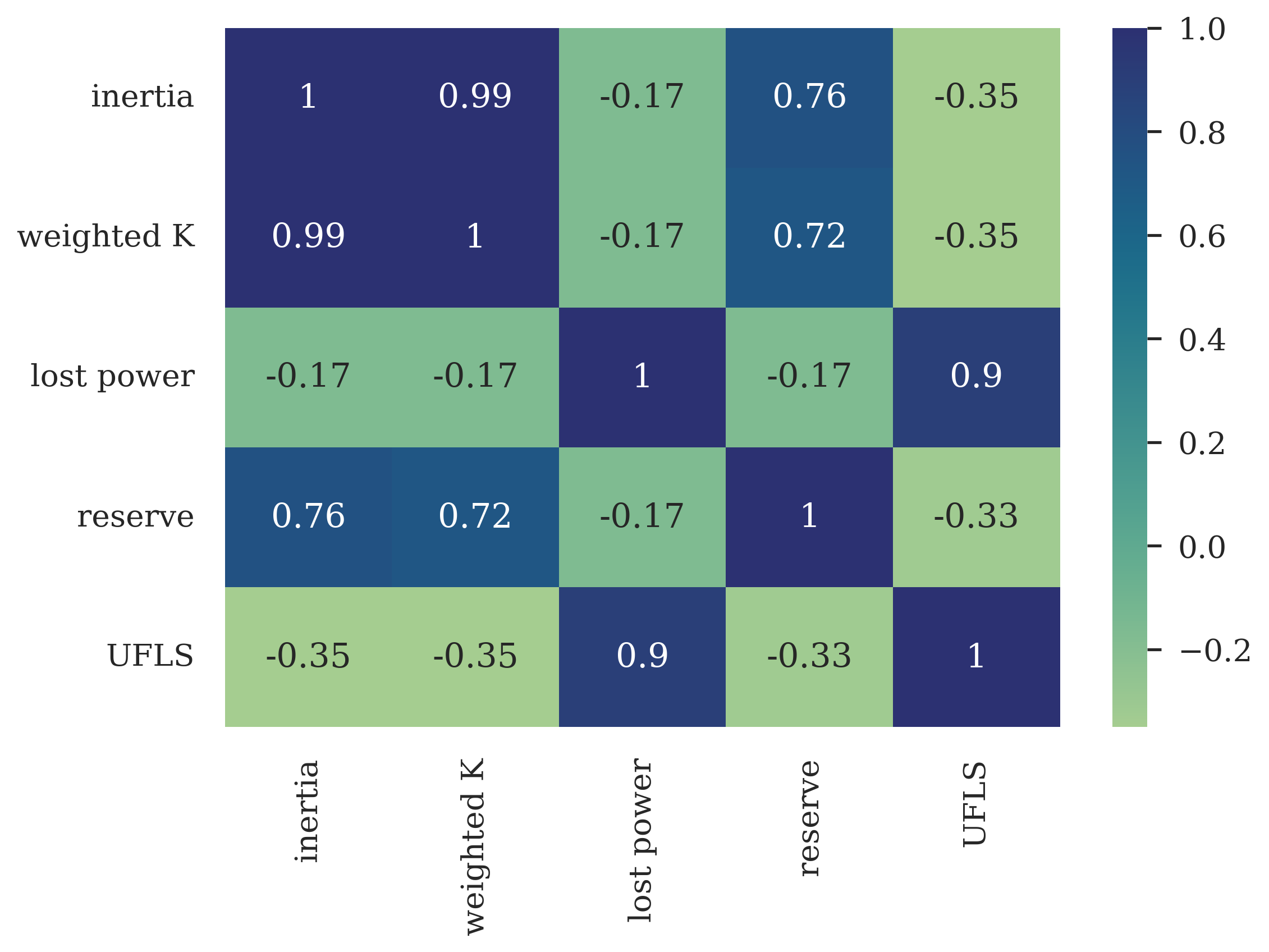}
\caption{Pearson correlation between inertia, weighted $K$, lost power, power reserve, and the amount of \gls{UFLS}.}
\label{correlation}\end{figure}

First, let's look at the correlations between the features and the labels. In \cref{correlation} the Pearson correlation between available inertia, weighted $K$, lost power, power reserve, and the amount of \gls{UFLS} is shown on a heatmap.

\begin{figure}[!htbp]
\centering
\includegraphics[width=1\linewidth]{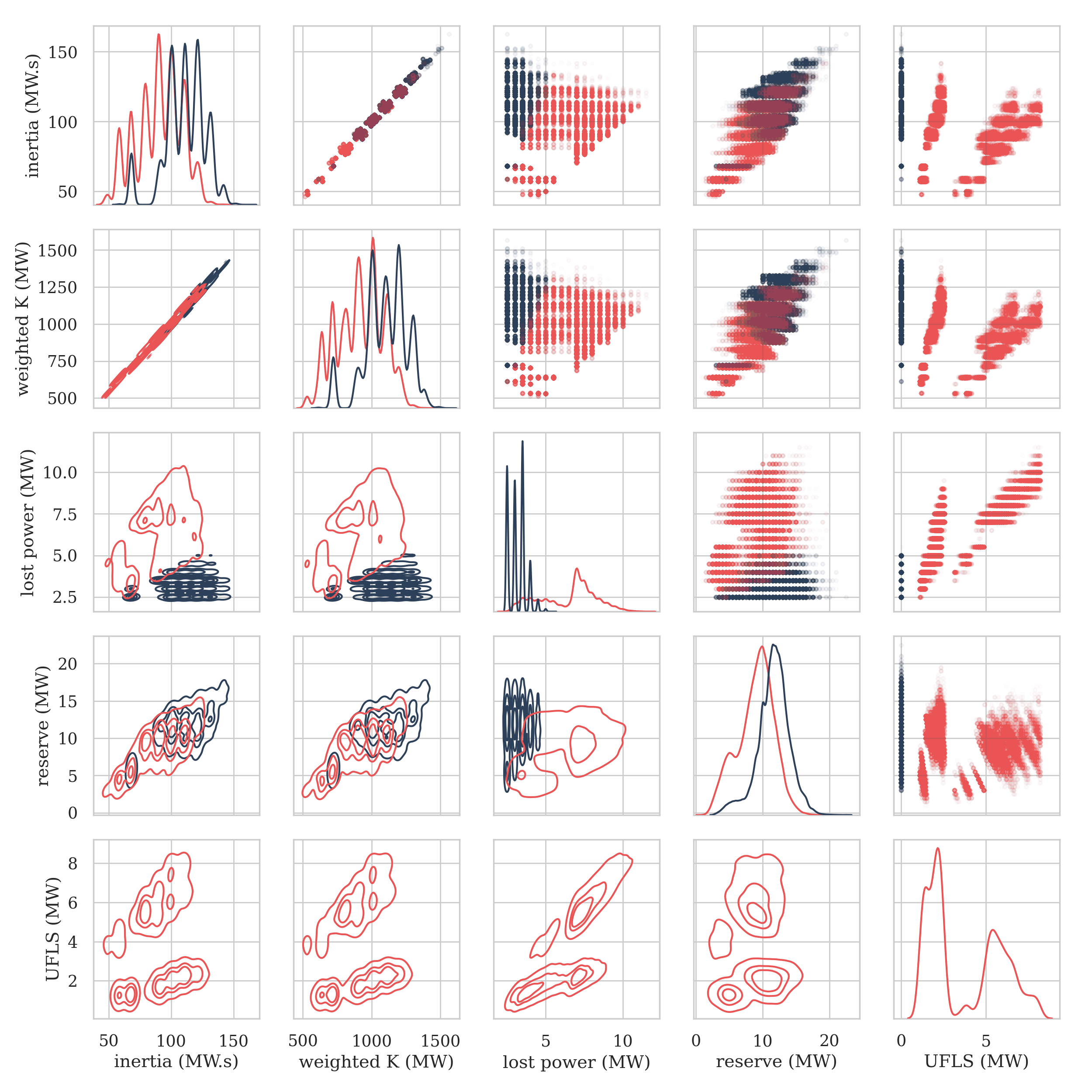}
\caption{\gls{KDE} and scatter plot of the features and the labels.}
\label{pairgrid}\end{figure}

\begin{figure}[!htbp]
\centering
\includegraphics[width=\linewidth]{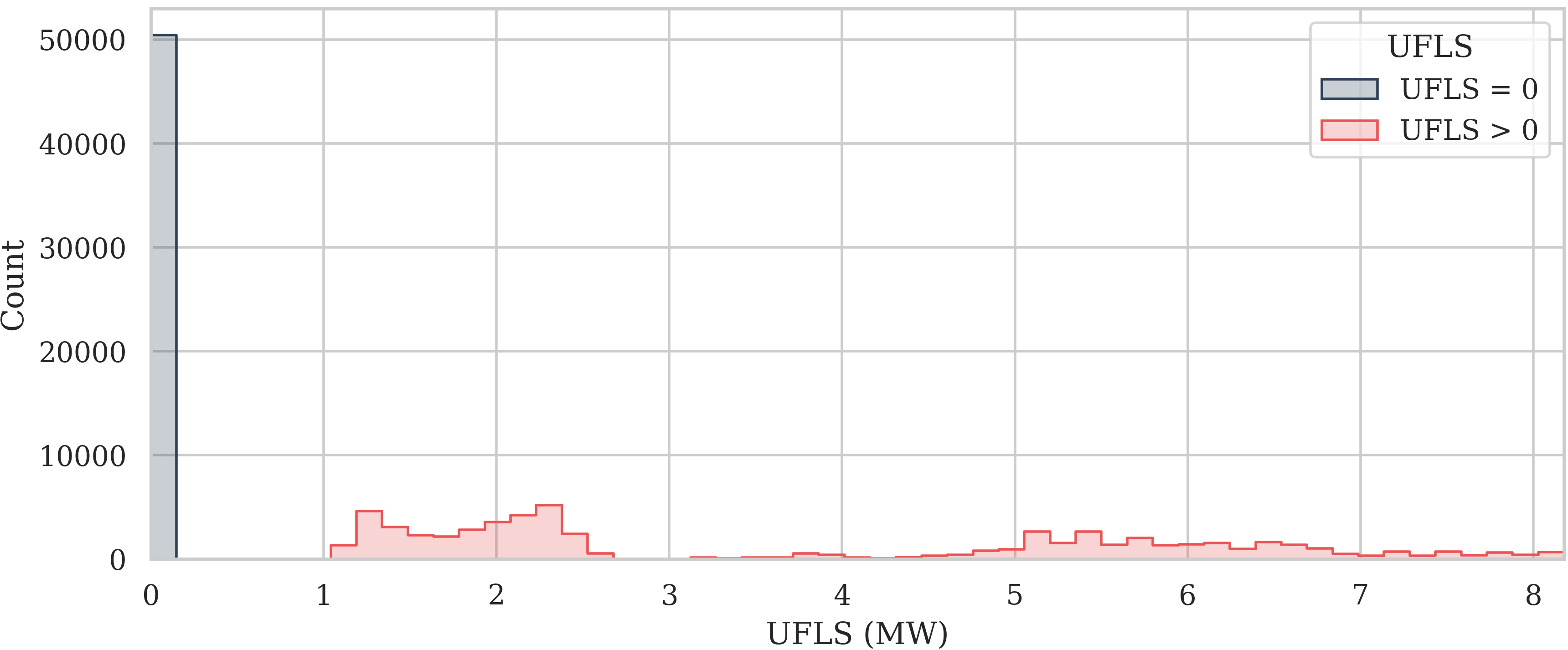}
\caption{Histogram of \gls{UFLS} amount.}
\label{histo_ufls}\end{figure}

In \cref{pairgrid} the \gls{KDE} plot of inertia, weighted $K$, lost power, power reserve, and the amount of \gls{UFLS} is depicted in the lower part of the figure. In the upper part of the figure, the scatter plot of the same quantities is depicted. The outages that don't lead to any \gls{UFLS} are shown in blue, and outages with positive \gls{UFLS} are shown in red.
\Cref{pairgrid} gives a good insight into the distribution of the data and how smooth the data is. This figure clearly shows the complexity of the problem at hand. As the final purpose is to use the estimation of \gls{UFLS} in the operational planning process, it's important to estimate the blue incidents in \cref{pairgrid} as exactly zero, and not a small number.
Although the general relation between the features shown in \cref{pairgrid} and the amount of \gls{UFLS} is complex and non-linear, some trades can be spotted. It seems that the incidents with no \gls{UFLS} (in blue), and the incidents with some \gls{UFLS} (in red) cannot be easily distinguished with only one feature. The combination of all features will distinguish blue and red dots with better accuracy. That's another reason to use methods like logistic regression for splitting the nodes, rather than decision trees that rely on one feature to apply the splits.
In \cref{histo_ufls} a histogram of \gls{UFLS} is presented.
Both of the methods that were introduced in the methodology (Tobit model and proposed regression tree) are applied to the dataset in order to estimate \gls{UFLS}.

\subsection{Learning Process}\label{sec:learning_res}

To train and evaluate the models, the dataset from \cref{sec:data_analysis} is divided randomly into a training dataset (80\% of the data) and a test dataset (20\% of the data). The learning process is done using the training dataset, and the evaluation is done using the test dataset.

\subsubsection{With Regression Tree}

\begin{figure}[!htbp]
\centering
\includegraphics[width=\linewidth]{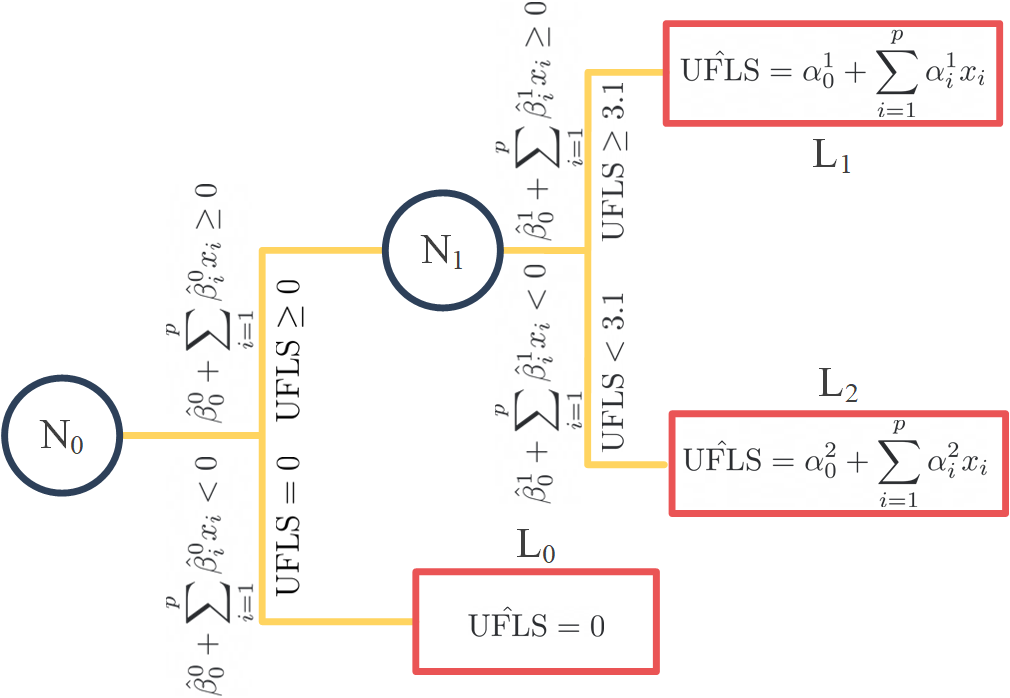}
\caption{Proposed regression tree with minimum overall \gls{MAE}.}
\label{fig:best_tree}\end{figure}

A grid search is performed to find the optimum tree structure. Different \gls{UFLS} thresholds for splitting are tried in a loop, starting from zero and with 0.1 MW steps, and the one that leads to overall minimum \gls{MAE} is chosen. Looking at the distribution of the amount of \gls{UFLS} in \cref{pairgrid} three groups of data can be distinguished: incidents with zero \gls{UFLS}, incidents with small \gls{UFLS} (between 0 to 4 MW), and incidents with big \gls{UFLS} (between 4 to 8 MW). Considering this observation and after performing a grid search to find $c$, the tree structure shown in \cref{fig:best_tree} achieves small \gls{MAE} while being simple.
On the node N$_0$ the data is classified into positive \gls{UFLS} and zero \gls{UFLS}. The split containing zeros doesn't need any further classification as all of them are equal to zero. On the node N$_1$ the remaining points are classified into \gls{UFLS} bigger than 3.1 MW and smaller than 3.1 MW. The estimated amount of \gls{UFLS} is presented by its corresponding linear regression on each leaf. The obtained data of this tree structure is presented in \cref{tab:data_tree}. The scores that are shown in this table are the result of applying the model trained by the training dataset, on the test dataset.
\begin{table*}[!htbp]
\small
    \centering
    \caption{The data of the trained tree.}
    \label{tab:data_tree}
    \begin{tabular}{c|ccccc|c}
         & intercept &$\times\mathcal{H}_g$ & $\times\mathcal{K}_g$ & $\times P_g$ & $\times\mathcal{R}_g$ & score \\\toprule
        N$_0$ & $\beta_0^0$=1 &$\beta_1^0$=0.516 &$\beta_2^0$=-0.158 & $\beta_3^0$=28.957 & $\beta_4^0$=-1.176 & acc=98.8\%  \\
        N$_1$ & $\beta_0^1$=1 &$\beta_1^1$=-0.003 &$\beta_2^1$=0.004 & $\beta_3^1$=-0.726 & $\beta_4^1$=0.048 & acc=94.7\% \\
        L$_0$ & $\alpha_0^0$=0 &$\alpha_1^0$=0 &$\alpha_2^0$=0 & $\alpha_3^0$=0 & $\alpha_4^0$=0 & \gls{MAE}=0 \\
        L$_1$ & $\alpha_0^1$=-0.104 &$\alpha_1^1$=0.019 &$\alpha_2^1$=0.0001 & $\alpha_3^1$=0.106 & $\alpha_4^1$=-0.057 & \gls{MAE}=0.037 MW \\
        L$_2$ & $\alpha_0^2$=0.006 &$\alpha_1^2$=0.026 &$\alpha_2^2$=-0.002 & $\alpha_3^2$=0.870 & $\alpha_4^2$=-0.176 & \gls{MAE}=0.272 MW \\
    \end{tabular}
\end{table*}

\begin{figure}[!htbp]
\centering
\includegraphics[width=\linewidth]{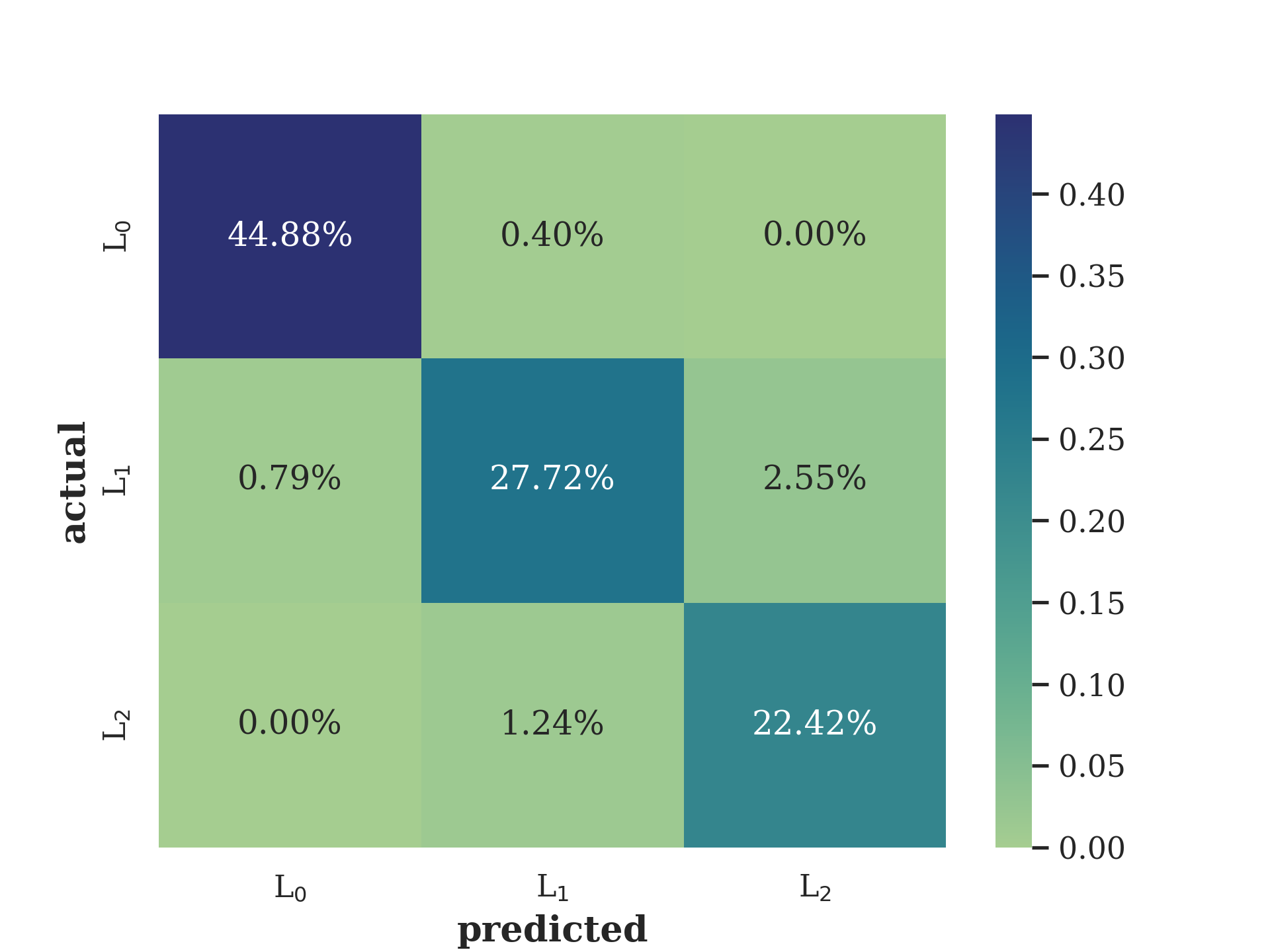}
\caption{The classification confusion matrix for each class.}
\label{fig:classi_heatmap}\end{figure}

The final \gls{MAE} of this process will be partly due to the classification errors on the nodes and the regression error on the leaves. In \cref{fig:classi_heatmap} the classification error on different leaves is presented.
The diagonal squares show the percentage of true positive classifications for each class. For example, $95.02\%$ of the whole data has been classified correctly (sum of the diagonal squares). $0.79\%$ of the samples on L$_1$ are incorrectly classified as L$_0$. $0.40\%$ of L$_0$ samples are incorrectly classified as L$_1$, and so on.

\begin{figure}[!htbp]
\centering
\includegraphics[width=\linewidth]{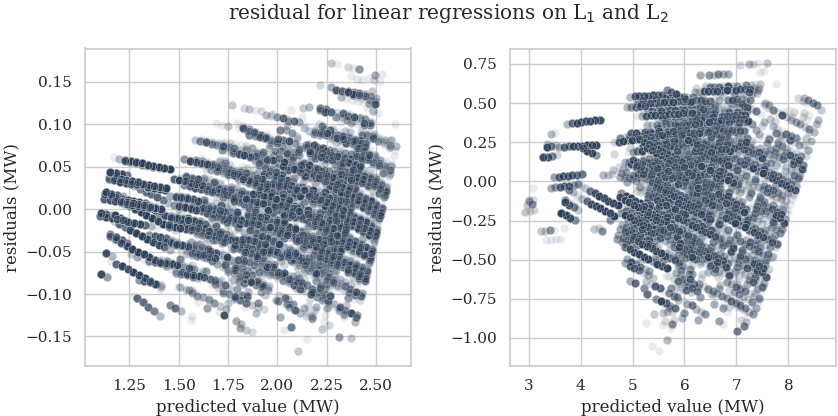}
\caption{The residual for the regression applied on L$_1$ and L$_2$.}
\label{fig:UFLS_reg}\end{figure}

The \gls{UFLS} for all the samples that are assigned to L$_0$ is estimated as zero and for L$_1$ and L$_2$ linear regression is applied. The residuals (predicted value minus observed labels) of the estimation on the test dataset are shown in \cref{fig:UFLS_reg}.
Considering the complexity of the problem at hand, and being limited to using linear models, the accuracy is acceptable. The prediction error for the samples in L$_1$ is always less than 0.15 MW and for L$_2$ is always less than 1 MW. Note that the samples on L$_2$ are bigger than the samples on L$_1$.

\begin{figure}[!htbp]
\centering
\includegraphics[width=\linewidth]{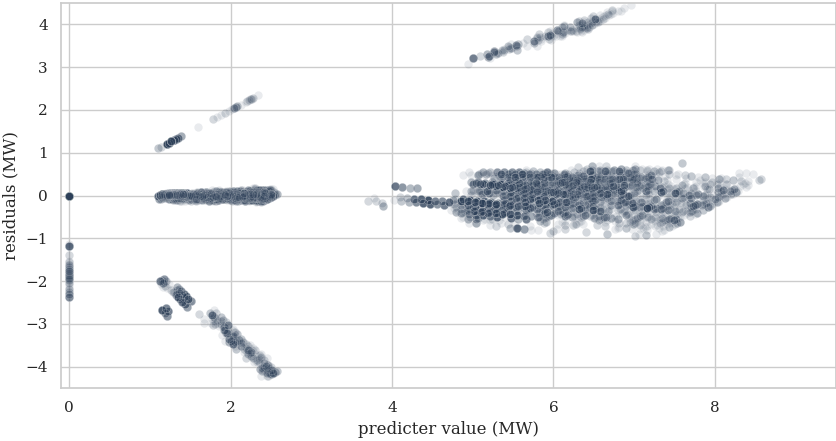}
\caption{The residual for the suggested regression tree.}
\label{fig:res_all}\end{figure}

Now considering all of the classifications and regression applied in the suggested tree structure, it's possible to look at the residuals for the whole process, shown in \cref{fig:res_all}.
Other than regression errors, errors due to misclassification are evident. The bigger residuals are because of the misclassification. That's why more complicated tree structures wouldn't improve the overall accuracy in this case. Note that although the estimation error might be high for some incidents, it doesn't endanger the stability of the system as the \gls{UFLS} scheme will ensure the stability of the system. It's expected that the benefits from correct estimations will outweigh the downsides of errors. The \gls{MAE} of the whole process on the test dataset is 0.213 MW. The trained tree can be represented as \gls{MILP} with the addition of 18 new constraints, 3 continuous variables, and 3 binary variables for every observation.

\subsubsection{With Tobit Model}

\begin{figure}[!htbp]
\centering
\includegraphics[width=\linewidth]{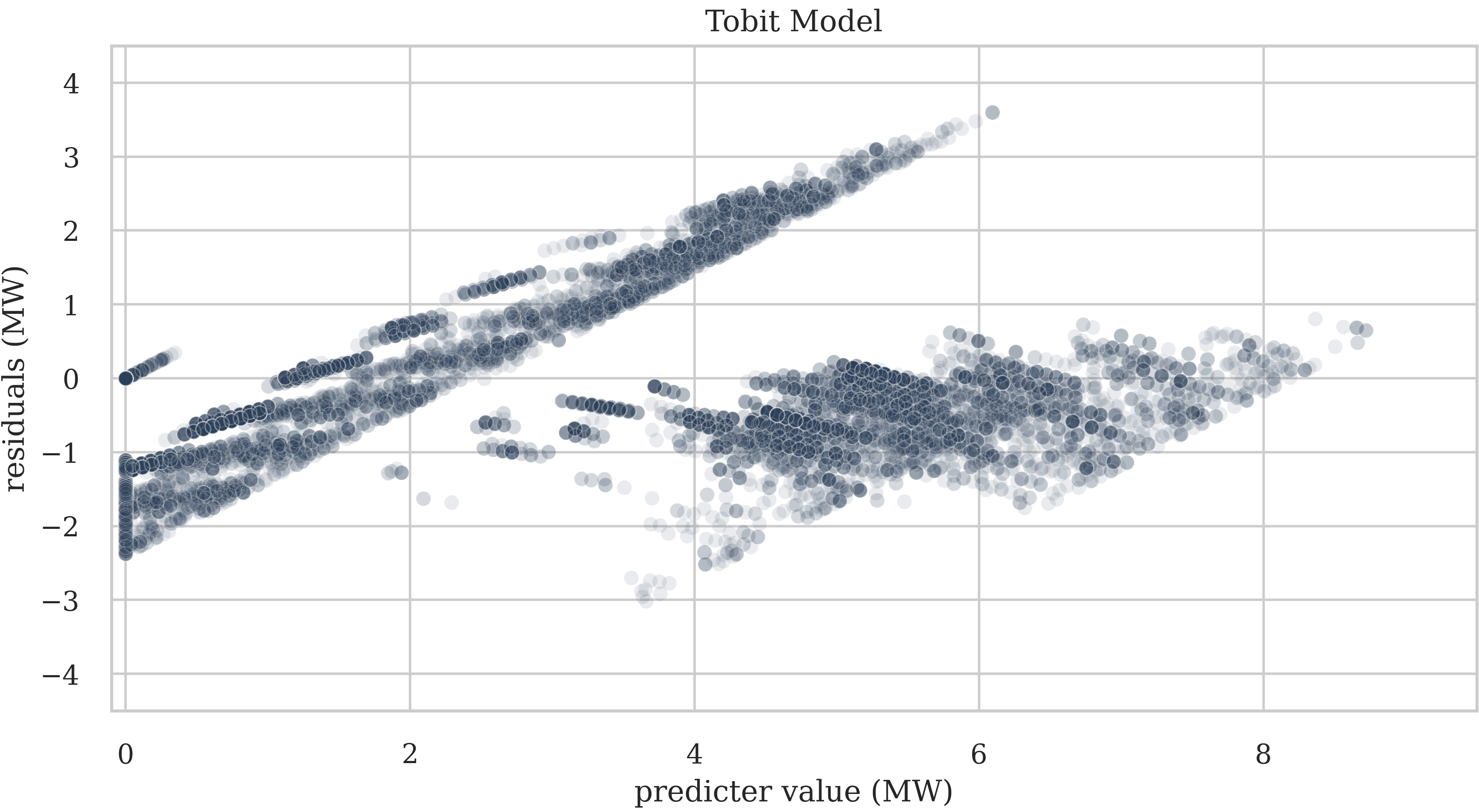}
\caption{Residuals of the test dataset. The training dataset is trained with Tobit model.}
\label{fig:tobit_res}\end{figure}

The training dataset is trained with the Tobit model (\cref{sec:tobit_metho}). Then it's applied to the test dataset. The residuals are shown in \cref{fig:tobit_res}.
The model has successfully distinguished most of the zero \gls{UFLS} incidents and pushed them to the negative side so they will be equal to zero in the model. As the model tries to fit all of the positive \gls{UFLS} incidents with one line, the error is high for some incidents. The overall \gls{MAE} of the model on the test dataset is 0.4967 MW. The advantage of this model is being easy to implement as \gls{MILP}. Here is the trained Tobit model,
\begin{equation}\label{eq:tobit_res}
    \hat{\text{UFLS}}=\begin{cases}
    -0.702-0.027\mathcal{H}_g-0.001\mathcal{K}_g\\+1.382P_g-0.132\mathcal{R}_g & \text{if it's}>0 \\
    0 & \text{otherwise}
  \end{cases}
\end{equation}
According to the \cref{eq:tobit_lin} the term in \cref{eq:tobit_res} can be represented as \gls{MILP} with the introduction of 2 new binaries and 3 constraints for each observation. 

\section{Conclusion}\label{sec:conc}

In this paper, a \gls{ML}-based approach for estimating \gls{UFLS} in power systems is presented. By leveraging a carefully generated dataset and applying two suggested \gls{ML} algorithms, the proposed regression tree, and the Tobit model, the relationship between relevant features and \gls{UFLS} labels is learned. The trained model demonstrated accurate and effective \gls{UFLS} estimation, providing valuable insights for operational planning, that will lead to frequency response improvement, reserve allocation optimization, and cost reduction. Applying the methodology to the La Palma island power system showcased its practicality and reliability, highlighting the potential for integrating \gls{UFLS} estimation into the scheduling optimization problem. While the \gls{MILP} representation of the Tobit model is computationally simpler, the accuracy of the suggested binary tree structure is superior. Future research avenues may focus on integrating this methodology into the actual operational planning problems like \gls{UC} and \gls{ED}.

Exploring additional features, investigating alternative \gls{ML} algorithms, and considering the impact of varying system configurations can further enhance the accuracy and applicability of \gls{UFLS} estimation. To follow up on the findings of this paper, the proposed models should be implemented in operational planning problems, like \gls{UC}, to further prove its benefits.

\section*{Acknowledgment}

This research has been funded by grant PID2022-141765OB-I00 funded by MCIN/AEI/ 10.13039/501100011033 and by “ERDF A way of making Europe”

\bibliographystyle{IEEEtran}
\bibliography{cas-refs}\label{references}

\end{document}